\documentclass[12pt, onecolumn, final]{IEEEtran}
\usepackage[latin1]{inputenc}
\usepackage{graphicx}
\graphicspath{{./Figures/}}
\usepackage{amssymb}
\usepackage{array}
\usepackage{color}
\usepackage{subfigure}
\usepackage[normalem]{ulem} 

\newcommand{\mat}[1]{\mathbf{#1}}

\newcommand{\refeq}[1]{(\ref{#1})}
\newcommand{\reffig}[1]{Fig.\ref{#1}}
\newcommand{\refsec}[1]{Section \ref{#1}}
\newcommand{\reftab}[1]{Tab.\ref{#1}}

\begin{document}
\title{Macro and Micro Diversity Behaviors of Practical Dynamic Decode and Forward Relaying schemes.}
\author{\IEEEauthorblockN{Mélanie Plainchault\IEEEauthorrefmark{1}, Nicolas Gresset\IEEEauthorrefmark{1}, Ghaya Rekaya-Ben Othman\IEEEauthorrefmark{2}}\\
\IEEEauthorblockA{\IEEEauthorrefmark{1}Mitsubishi Electric R\&D Centre Europe, France}\\
\IEEEauthorblockA{\IEEEauthorrefmark{2}Télécom ParisTech, France}
}
\maketitle

\begin{abstract}
In this paper, we propose a practical implementation of the Dynamic Decode and Forward (DDF) protocol based on rateless codes and HARQ.

We define the macro diversity order of a transmission from several intermittent sources to a single destination.
Considering finite symbol alphabet used  by the different sources, upper bounds on the achievable macro diversity order are derived. We analyse the diversity behavior of several relaying schemes for the DDF protocol, and we propose the Patching technique to increase both the macro and the micro diversity orders.
The coverage gain for the open-loop transmission case and the spectral efficiency gain for the closed loop transmission case are illustrated by simulation results.
\end{abstract}

\section{Introduction}
Relays have been introduced in wireless communication systems in order to improve the transmission's quality. Indeed, the combination of a source and a relay forms a virtual Multiple Inputs Multiple Outputs (MIMO) scheme providing diversity and robustness to fadings.
Two main relaying protocols can be distinguished in the literature: the Amplify and Forward (AF) protocol and the Decode and Forward (DF) protocol. For the AF protocol, the relay transmits an amplified copy of the previously received signal. The main drawback of this protocol is the noise amplification. Using a DF protocol, the relay sends a decoded version of the previously received symbol. In this case, the relay can introduce decoding errors which can lead to error propagation.
Azarian et al. proposed in \cite{azarian2005adm} the Dynamic Decode and Forward (DDF) protocol (or Sequential Decode and Forward in \cite{katz2009cssonr}) to avoid both drawbacks. In the DDF protocol, the relay switches into a transmission mode only after having correctly decoded the message from previously received signals. As the source-relay link is perturbed by a random short-term fading coefficient, the instant of correct decoding at the relay is unknown or, in other words, dynamic.
The DDF protocol was widely studied in the litterature according to theoretical metrics: diversity multiplexing trade-off in \cite{murugan2006clc, elia2007aus} and capacity region in \cite{katz2009cssonr}. 

In this work, we focus on the practical implementation of the DDF protocol using error correcting codes and space-time codes.
We study open-loop transmissions, evaluated according to the outage probability in which no feedback is allowed between the destination and the source.
We also consider closed-loop transmissions, evaluated according to the system spectral efficiency in which feedback is allowed.
Our proposed protocol is taking benefit from rateless codes \cite{castura2007rcwrc} and Hybrid Automatic Repeat reQuest (HARQ) \cite{byun2008cha}.

The well-known diversity order achievable by a cooperative scheme is called \textit{micro diversity} which is generated by short term fading coefficients. Another kind of diversity, called \textit{macro diversity}, can be obtained by observing several long term Signal-to-Noise Ratios (SNR), \cite{bernhardt1987macrodiv, Tang2001mmd,gresset2010mmp}.
The main contribution of this paper is to define a macro diversity order for DDF protocols and to show the improvement provided by the exploitation of the macro diversity. Moreover, we recall and expand the so-called Patching technique, we have proposed in \cite{plainchault2010pdstbc}, in order to improve the achievable macro and micro diversity orders.

In \refsec{Part:SysMod}, the considered system model and the assumptions are described. Figures of merit are presented in \refsec{Part:Metrics}.
In \refsec{Part:MacroDiv}, the macro diversity order achievable by a transmission is defined for DDF transmission in a network containing multiple relays. The proposed DDF protocols are then designed and improved for the single-relay case.
The micro diversity order is also improved in \refsec{Part:MicroDiv} by using the Patching technique and/or space-time codes. Finally, simulation results are given in \refsec{Part:Perf}.

\section{System model and cooperative protocol}\label{Part:SysMod}
\begin{figure}
	\centering
	\includegraphics[width=0.99\columnwidth]{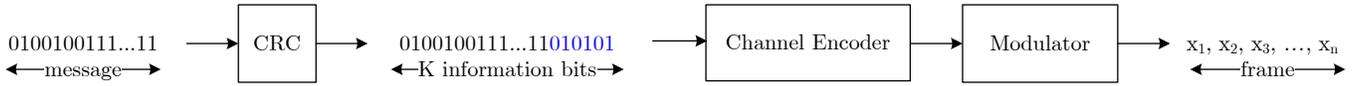}
	\caption{\label{fig:CodewordGen}Frame generation by the source.}
\end{figure}

We consider a wireless system comprising a source S, an in band reception/transmission half-duplex relay R, each carrying a single transmission antenna, and a destination D carrying $N_r$ reception antennas. A relay is said to be half-duplex when it cannot transmit and receive at the same time and is said to be an in band reception/transmission relay when it receives and transmits on the same physical resource as the source \cite{sahin2009}.
%
We further assume in some cases that the source is \textit{ relay-unaware}.
This assumption avoids the need of pilots or control signals dedicated to relaying on the source-relay or relay-source links and allows autonomous relay configuration.
%
We propose a practical design for DDF protocols based on HARQ transmission scheme \cite{byun2008cha}\cite{castura2007rcwrc}. The source generates a frame to be transmitted to the destination according to the process illustrated in \reffig{fig:CodewordGen}.

The $K$ information bits contain Cyclic Redundancy Check (CRC) bits allowing for the destination to check the message correct decoding.  We assume that the number of added CRC bits is negligible when compared to the number of message bits  and neglect the false positive CRC events.
These $K$ information bits are then encoded by a channel encoder, for instance with a rate matching algorithm, in order to generate $N_{max}$ sub-frames of coded bits. The concatenation of these sub-frames is called a frame, and the resulting channel coding rate is denoted $R_c$. The bits of the concatenated sub-frames are then modulated to form symbols from a finite alphabet, such as a $2^{m_S}$-QAM. 
The $i$-th sub-frame is composed of $T_i$ symbols or $T_im_S$ coded bits. All sub-frames of symbols are separated in time in order to allow a processing delay at the receiver side as for classical transmission with HARQ.
One codeword transmission is limited by the transmission of $\sum_{i=1}^{N_{max}}T_i$ symbols.

\begin{figure}
	\centering
	\includegraphics[width=0.99\columnwidth]{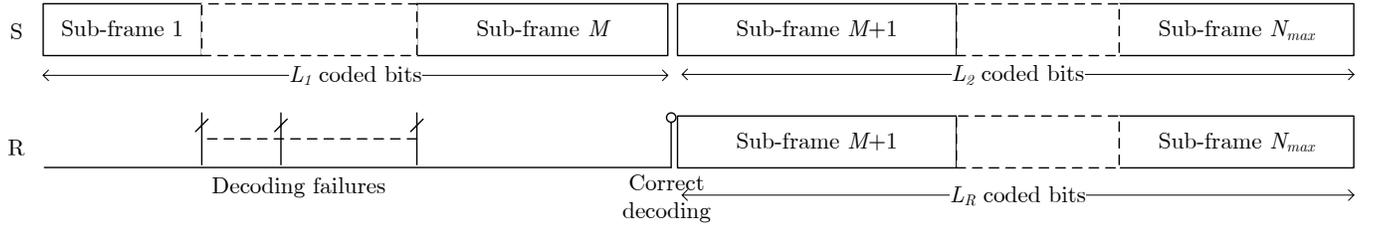}
	\caption{\label{fig:FrameDetail} A codeword, or frame, is segmented into $N_{max}$ sub-frames, the $i$-th sub-frame is composed of $T_i$ time-slots. The relay transmits, on the same physical resource as the source, after correctly decoding the sent message.}
\end{figure}

By definition, we assume that during the first phase of the DDF protocol, the relay tries to decode the message intended to the destination after each sub-frame transmitted by the source.
As soon as the relay has correctly decoded this message, it switches into a transmission mode, which defines the second phase of the DDF protocol. 
Assuming perfect synchronization, the relay then transmits a signal on the same physical resource used by the source, according to one of the proposed relaying scheme such as Monostream DDF, Distributed Alamouti DDF or DDF with Patching, further detailed in Sections \ref{Part:MacroDiv} and \ref{Part:MicroDiv}. During the second phase, if needed, the relay transmits dedicated control and pilot signals to the destination.
We denote $L_{1}$ and $L_2$ the number of coded bits transmitted by the source during the first and second phase respectively, while the relay transmits $L_R$ bits using a $2^{m_R}$-QAM.
The frame structure is presented in \reffig{fig:FrameDetail}.

The destination attempts to decode the message after each received sub-frame according to the phase of the DDF protocol.
After a correct decoding, the destination stops listening, and sends an acknowledgment (ACK) to the source or not, depending on the transmission mode. As further described in Section \ref{Part:Metrics}, the source stops or not the frame transmission, depending on the transmission mode.

We consider quasi-static Rayleigh fading channels, i.e., the fading coefficients remain constant during at least the transmission of a frame and are independent from one frame to another. The fading vectors $\mat{h_{SD}}$ between the source and the destination, $\mat{h_{RD}}$ between the relay and the destination,  are vectors of $N_r$ complex-Gaussian distributed, zero-mean and unit variance random values.
A complex-Gaussian noise of zero-mean and variance $N_0$ per real dimension is added at each reception antenna.
The received power at the destination side from transmitter $R_j$ is denoted $P_{R_j}$. It represents the transmit power affected by the path loss, the antenna gains and the shadowing. More precisely, $P_{R_0}=P_S$ is the received power at the destination side from the source, and $P_R $ is the received power at the destination from the relay.
In the following, the values of the long term signal to noise ratios are denoted $SNR_{SR}$, $SNR_{SD}=P_S/(2N_0)$ and $SNR_{RD}=P_R/(2N_0)$ for the  source-relay, source-destination and relay-destination links, respectively.

\section{Characterization of the efficiency of a DDF scheme}\label{Part:Metrics}

In this Section, we define the figures of merit for open-loop and closed-loop transmissions with HARQ.

When using the DDF protocol detailed in Section \ref{Part:SysMod}, the mutual information observed after the $n$-th sub-frame by assuming that the relay transmits from the beginning of the $M$-th sub-frame is defined by
\begin{equation}\label{eq:InfoMutDDFsimplified}
 I_\textrm{\small{D}}^{(n, M)} = I_{\textrm{\small{S}}\rightarrow \textrm{\small{D}}}^{(n,M)}+ I_{\textrm{\small{\{S,R\}}}\rightarrow\textrm{\small{D}}}^{(n,M)} \quad \textrm{with } 1<M \leq n
\end{equation}
where $I_{\textrm{\small{S}}\rightarrow \textrm{\small{D}}}^{(n,M)} = \frac{I_{\textrm{\small{S}}\rightarrow \textrm{\small{D}}}\sum_{i=1}^{M-1}T_i}{\sum_{i=1}^{n}T_i}$ is the mutual information $I_{\textrm{\small{S}}\rightarrow \textrm{\small{D}}}$ observed when only the source transmits, weighted by the ratio between the length of the first phase and the total codeword length. It depends on the spectral efficiency $m_{S}$ of the QAM modulation used by the source, the $SNR_{SD}$ and the fading coefficient $\mat{h}_{SD}$ between the source and the destination D.
The mutual information $I_{\textrm{\small{\{S,R\}}}\rightarrow \textrm{\small{D}}}^{(n,M)}= \frac{I_{\textrm{\small{\{S,R\}}}\rightarrow\textrm{\small{D}}}\sum_{i=M}^{n}T_i }{\sum_{i=1}^{n}T_i} $ is observed during the second phase of the DDF protocol and depends on $m_{S}$, $SNR_{SD}$, $\mat{h}_{SD}$ but also on the relay-destination link $SNR_{RD}$, $m_{R}$, $\mat{h}_{RD}$ and on the relaying scheme.
We denote the fact that the relay does not transmit during the $n$ sub-frames, i.e., when no correct decoding occurred at the relay, by $M=\varnothing$. Thus,
\begin{equation}
I_\textrm{\small{D}}^{(n,\varnothing)} = I_{\textrm{\small{S}}\rightarrow \textrm{\small{D}}}.
\end{equation}
Similarly, $I_{\textrm{\small{S}}\rightarrow \textrm{\small{R}}}$ is the mutual information observed between the source and the relay and depends on the spectral efficiency $m_{S}$ of the QAM modulation used by the source, the $SNR_{SR}$ and the fading coefficient between the source and the relay R.

The destination D is  in outage after receiving the $n$-th subframe knowing that the relay begins to transmit during sub-frame $M$, if the mutual information $I^{(n, M)}_\textrm{\small{D}}$ is lower than the data rate $R_{n}$ used by the source. Because the number of CRC bits is neglected when compared to the number of bits in the message, we define the data rate as follows:
\begin{equation}
R_{n} = \frac{K}{\sum_{i=1}^{n}T_i}.
\end{equation}
Thus, the outage probability observed at D after receiving the $n$-th sub-frame, knowing that the relay begins to transmit during the $M$-th sub-frame, is defined as:
\begin{equation}
 P_{out, \textrm{\small{D}}}^{(n, M)} = \textrm{Prob }(I^{(n, M)}_\textrm{\small{D}}<R_{n}).
\end{equation}
After averaging on the instant of correct decoding at the relay, the outage probability observed at D after receiving the $n$-th sub-frame is
\begin{equation}\label{eq:Pout_def}
P_{out, \textrm{\small{D}}}^{(n)} = \sum_{M=2}^{n}P_{out, \textrm{\small{D}}}^{(n, M)}P_{1^{st}, \textrm{\small{R}}}^{(M-1)}+P_{out, \textrm{\small{D}}}^{(n, \varnothing)}P_{out, \textrm{\small{R}}}^{(n-1)}
\end{equation}
where $P_{1^{st}, \textrm{\small{R}}}^{(M-1)}$ is the probability that the relay correctly decodes the message after receiving the $M-1$-th sub-frame  and not before, i.e,
\begin{equation}
P_{1^{st}, \textrm{\small{R}}}^{(M-1)} = \textrm{ Prob }\left ( R_{M-1} \leq  I_{\textrm{\small{S}}\rightarrow \textrm{\small{R}}} < R_{M-2} \right)
\end{equation}
and
\begin{equation}
P_{out, \textrm{\small{R}}}^{(n-1)} = 1- \sum_{i=1}^{n-1}P_{1^{st}, \textrm{\small{R}}}^{(i)}= \textrm{Prob} (I_{\textrm{\small{S}}\rightarrow \textrm{\small{R}}} < R_{n-1}).
\end{equation}
Similarly, $P_{1^{st}, \textrm{\small{D}}}^{(n, M)}$ is the probability that the destination correctly decodes the message after receiving the $n$-th  sub-frame and not before, knowing that the relay begins to transmit during the $M$-th sub-frame, and is defined by
\begin{equation}
P_{1^{st}, \textrm{\small{D}}}^{(n, M)} = \textrm{ Prob }\left ( R_{n} \leq I_\textrm{\small{D}}^{(n, M)} < R_{n-1} \right).
\end{equation}

\subsection{Figure of merit for open-loop transmissions}
A broadcast (or multicast) system is a practical example in which open-loop transmission occurs.
In such a system, a source sends a common message to several destinations. Thus, it cannot adapt its transmission  neither to each destination nor to any relay that would be in the system. Moreover, this illustrates a particular case in which the assumption of relay-unaware source is particularly relevant.
Thus, the modulation, the coding rate, and the frame length are chosen to meet a required Quality of Service (QoS) in the worst wireless link conditions, which defines the system coverage.
For a given data rate, the coverage is improved by decreasing the outage probability which is directly linked to the QoS.
In the following, we will consider $P_{out, \textrm{\small{D}}}^{(N_{max})}$ as the figure of merit for open loop transmissions.

\subsection{Figure of merit for closed-loop transmissions with HARQ}
When HARQ with incremental redundancy is considered, the overall spectral efficiency is improved by allowing the destination to acknowledge the correct decoding of the message after each received sub-frame.
The spectral efficiency $\mathcal{S}_{HARQ}$ of our practical DDF scheme can be expressed as
\begin{equation}\label{eq:SpecEffHARQ}
 \mathcal{S}_{HARQ} = \sum_{n=1}^{N_{max}} R_{n} \left[ P_{out, \textrm{\small{R}}}^{(n-1)}P_{1^{st}, \textrm{\small{D}}}^{(n, \varnothing)}+\sum_{m=1}^{n-1}P_{1^{st}, \textrm{\small{R}}}^{(m)}P_{1^{st}, \textrm{\small{D}}}^{(n, m)} \right].
\end{equation}

\section{Macro diversity order of DDF relaying schemes}\label{Part:MacroDiv}
In a system comprising several highly separated sources transmitting the same signal, the SNR observed at the destination is improved by a higher received power and a diversity on the path gain when compared to the single source case. The path gain encompasses the path loss, the random shadowing and the source and destination antenna diagram gain.
Our target is to improve the system for all possible destination positions. By considering a random variation of the destination position, the SNR becomes a random variable and the performance averaged on its probability density function is improved by increasing the macro diversity order \cite{Tang2001mmd}, equal to the number of sources. If one SNR link is low, other links are observed to support the transmission.

Let us consider a system comprising a destination observing a significant path gain from both a source and a relay.
When the relay never transmits, the macro diversity order observed at the destination is equal to one.
When both the source and the relay transmit the whole frame, the macro diversity order is two.
However, the random activation of the relay in the DDF protocol makes the number of sources vary through time, and the relay only sends a fraction of the codeword. A definition of the achievable macro diversity order is needed in this case. 

The definition of a macro diversity order achievable by a system comprising several relays is given in \refsec{sPart:MacroDef}. In \refsec{sPart:MacroMono}, we describe the equivalent long term SNR channel experienced by a Monostream DDF transmission, and we derive a macro diversity order upper bound. 
Solutions for improving the macro diversity order are then presented in \refsec{sPart:MacroImp} for the single relay case.

\subsection{Definition of the macro diversity order}\label{sPart:MacroDef}

\textit{Definition}
Given a figure of merit $U$ function of $n$ long-term SNRs $(\rho_1, \cdots, \rho_n)$ and increasing according to each variable $\rho_i$ taken separately; given a target value $U_t$; the macro diversity order $d$ for the target value $U_t$ is defined by
\begin{equation}
d = \min_{\Omega \subset \{1:n\}} \left (\left|\Omega\right|\left|\lim_{\renewcommand{\arraystretch}{0.5}
\begin{array}{>{\scriptstyle}c}
\forall j\in\Omega, \rho_j \to 0 \\
\forall i\notin\Omega,\rho_i \to +\infty
\end{array}}
U(\rho_1, \cdots, \rho_n)< U_t \right.\right).
\end{equation}
Thus, the macro diversity order is defined as the minimal number of links to turn off so that the target $U_t$ is not longer asymptotically achievable through the remaining links.
Note that the figure of merit $U$ could be, for instance, the spectral efficiency or the probability that the transmission is not in outage.

By definition, the full macro diversity order is achievable if $d=n$.
Consequently, the system is full macro diversity for the target value $U_t$ if and only if
\begin{equation}
\forall j \in \{1,\cdots,n\},\lim_{\rho_j \to +\infty}U(0,\cdots,0,\rho_j,0,\cdots,0) \geq U_t
\end{equation}
which means that every single link asymptotically allows to achieve the target.

As a remark, it is straightforward to see that, as soon as the relay transmits, a Gaussian input system always achieve the full macro diversity order. However, we will show in the following how a discrete input system limit the macro diversity exploitation.

\subsection{Macro diversity behavior of a Monostream DDF transmission scheme}\label{sPart:MacroMono}
By extending the system model to the multi-relay case, a DDF transmission with a source (denoted $R_0$) and $n$ relays is composed of $n+1$ phases, the $i$-th phase containing $L_i$ bits.
\subsubsection{Equivalent SNR channel for Monostream DDF transmission scheme}

Let's consider the simplest relaying scheme called Monostream DDF \cite{gresset2010mmp}, and also described in \cite{castura2007rcwrc} whose extension to the multiple relay case is straightforward. During their transmission phases, the relays send the same symbols as the source on the same frequency and time resource. This relaying scheme does not need dedicated pilots between the relays and the destination whose decoder complexity is kept low, and allows the source to be relay-unaware.
During a time-slot $k$ of the $i$-th phase, the destination receives $\mat y_{i,k}$ given by
\begin{eqnarray}
\mat y_{i,k} &=& \left(\sum_{j=0}^{i-1}\sqrt{P_{R_j}}\mat{h}_{{R_j}D}\right) x_{i,k} + \mat b_{i,k} \label{eq:SignalPhasei}
\end{eqnarray}
where $ x_{i,k}$ is the symbol sent during the k-th time-slot of the i-th phase, and $\mat b_{i,k}$ is the complex Gaussian noise vector of zero mean and variance $N_0$ per real dimension.

\begin{figure}
	\centering
	\includegraphics[width=0.7\columnwidth]{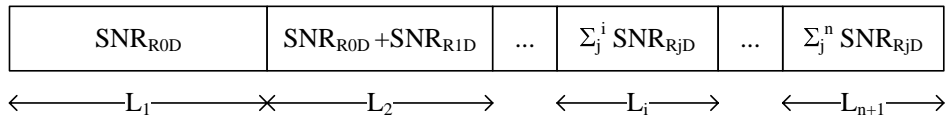}
	\caption{\label{fig:nSNRchannel}Long-term SNR channel of the Monostream DDF with n relays.}
\end{figure}

In order to decode the message, the destination computes soft bits from the likelihood ratios (LR) of each received symbol. The LR of the symbol sent during the $k$-th time-slot of the $i$-th phase is
\begin{eqnarray}
\frac{p(\mat y_{i,k}\mid x'_{i,k})}{p(\mat y_{i,k}\mid x_{i,k})}& = & exp \left (- \left|  \sum_{j=0}^{i-1}\sqrt{SNR_{R_jD}}\mat{h}_{{R_j}D}  \right| ^2(\mid x_{i,k}-x_{i,k}'\mid^2 + 2R_e(\mat b_{i,k}^*(x'_{i,k}-x_{i,k})) )\right ).
\label{eq:LLRphasei}
\end{eqnarray}
The $L_i$ soft bits all are a function of the same realization of a complex Gaussian fading law of zero mean and variance $\sum_{j=0}^{i-1}SNR_{{R_j}D}$.
Consequently, regarding the whole codeword, a transmission with Monostream DDF protocol follows the long term SNR channel presented on Fig.\ref{fig:nSNRchannel}. 
The equivalent long term SNR channel is composed of $n+ 1$ blocks, each one being characterized by a sum of long term SNR random variables.
The equivalent random SNR of the $i+1$ -th channel block is a coherent combination of the equivalent random SNR of the $i $-th channel block and an independent random variable $SNR_{R_{i+1}D}$. Thus, the equivalent block SNR channel is a Matryoshka channel, as defined in \cite{kraidy2007}, and denoted $\mathcal {M}((n+1,\cdots, 1),(L_{n+1},\cdots, L_1))$, where $(n+1,\cdots, 1)$ is the diversity order of each block sorted in a decreasing order and $(L_{n+1},\cdots, L_1)$ the number of bits in each block. 

\subsubsection{Macro diversity behavior of a Matryoshka SNR channel}
When considering the long-term SNRs as random variables, the diversity bound for discrete input block fading Matryoshka channel, derived in \cite{kraidy2007}, can be direclty applied. Consequently, the macro diversity observed after decoding a rate-$R_c$ linear code transmitted
over a $\mathcal{M}(\mathcal{D},\mathcal{L})$ long-term SNR channel is upper-bounded
by $\delta_{\mathcal{M}}(\mathcal{D},\mathcal{L})=\mathcal{D}_i$ where $i$ is given by the following
inequality:
\begin{equation}\label{eq:Matryoshkabound}
\sum_{k=1}^{i-1}{L}_k<R_c\sum_{k=1}^{N}{L}_k\leq\sum_{k=1}^{i}{L}_k.
\end{equation}
Therefore, the macro diversity behavior is linked to the number of bits contained in each phase of the DDF transmission, which is a function of the relays random activation time, and the coding rate.

We have derived the bounds on the macro diversity order for the discrete input Monostream DDF protocol for fixed relay activation configurations.
Unfortunately, the probability that the relay does not decode the message before the end of transmission while the destination does is non-null. This means that, theoretically and in average, the full macro diversity is never achieved. When this probability becomes negligible, this artefact does not impact the average performance, and the analysis of the macro diversity order based on fixed configuration holds.

\subsection{Macro diversity order improvement for the single-relay case}\label{sPart:MacroImp}
By applying the bound on the macro diversity order to a transmission with Monostream DDF and a single relay, we observe that the full macro diversity order is observed only if the relay correctly decodes the message early enough to satisfy $K \leq L_2$. Furthermore, this implies that the coding rate cannot exceed $1/2$ to reach full macro diversity.
In the following, we propose solutions for dynamically adapting the system to the relay decoding time in order to take the best benefit from the macro diversity.

\subsubsection{Macro diversity improvement with modulation adaptation}
Considering only one relay in the system, the full macro diversity order is achievable if and only if the second phase of the protocol carries at least $K$ bits.
By allowing signalling between the source and the relay, the macro diversity order can be improved by adapting the modulation size used by the source and the relay during the second phase of the transmission.
Both nodes then compute the same $L_2 \geq K$ redundancy bits using the same encoder based on the remaining number of symbols to be sent until the end of the codeword transmission. These bits are then modulated and transmitted to the destination.
This so-called Monostream DDF with modulation adaptation can be extended to the multiple relays case by 
adapting the modulation of all transmitting nodes so that full macro diversity is achievable according to the number of remaining time-slots in the frame after correct decoding at the last relay.

\subsubsection{Macro diversity improvement with Patching technique}
Considering the single-relay case, a Patching technique proposed in \cite{gresset2010mmp} and \cite{plainchault2010pdstbc} allows to improve the achievable macro diversity order without any signalling between the source and relay. Thus it can be used by the relay even when the source is relay-unaware.
This technique is a combination of two steps. The first one is done at the relay by transmitting combination of symbols already sent in the first phase and symbols the source is going to send in the second phase. The second step is done at the destination combining signals received during these different time-slots in order to build an equivalent transmission scheme using higher cardinality modulation.


More precisely, during the k-th time-slot of the second phase, the relay transmits $z_{k} \in 2^{m_R}$QAM, a combination of symbols $(x_{1,i})$ with $(k-1)(m_R/m_S-1)+1 \leq i \leq k(m_R/m_S-1)$.
For example, the relay forms $z_{k}\in$16QAM  from two QPSK symbols $x_{1,k}$ and $x_{2,k}$ sent during the first and second phase of the DDF protocol, respectively.
The respective signals received by the destination during the first and second phase are
\begin{eqnarray}
 \mat{y}_{1,i} & = & x_{1,i}\sqrt{P_S}\mat{h}_{SD}+\mat{b}_{1,i},\quad \forall (k-1)(m_R/m_S-1)+1 \leq i \leq k(m_R/m_S-1)\\
 \mat{y}_{2,k}   & = & x_{2,k}\sqrt{P_S}\mat{h}_{SD}+z_k\sqrt{P_R}\mat{h}_{RD}+\mat{b}_{2,k}.
\end{eqnarray}
For the particular case where the source uses QPSK symbols, the construction of the $2^{m_R}$QAM symbol sent by the relay is defined by:
\begin{eqnarray}
 z_k &=& f(x_{1,i}, x_{2,k}), \quad (k-1)(m_R/m_S-1)+1 \leq i \leq k(m_R/m_S-1)\\
 z_k &=& \sum_{i=1}^{m_R/m_S-1}a_ix_{1,i}+a_{m_R/m_S}x_{2,k},\quad a_i = \sqrt{\frac{3}{2^{m_R}-1}}2^{i-1}.
\end{eqnarray}
This particular construction realizes a bijection between the vector of combined symbols and the resulting symbol $z_k$. This property avoids rate deficiency for the decoding step at the destination side.

The destination then combines the received signals of phase 1 and 2 according to the symbol construction at the relay.
\begin{eqnarray}
\mat{y}_k & = & f(\mat{y}_{1,i}, \mat{y}_{2,k}),\quad (k-1)(m_R/m_S-1)+1 \leq i \leq k(m_R/m_S-1)\\
	& = & \sum_{i=1}^{m_R/m_S-1}a_i\mat{y}_{1,i}+a_{m_R/m_S}\mat{y}_{2,k} \label{eq:Patching}\\
	& = & z_k(\sqrt{P_S}\mat{h}_{SD}+ a_{m_R/m_S}\sqrt{P_R}\mat{h}_{RD}) + \mat{b}
\end{eqnarray}
where the resulting noise $\mat{b}$ is a complex Gaussian noise of zero mean and variance $N_0$ per real dimension. This combination \refeq{eq:Patching}, or \textit{Patching} operation at the destination side, is done to rebuild the QAM symbol $z_k$ making the decoding easier. The Patching operation generates a Matryoshka SNR channel whose block of highest diversity order contains more bits than the block obtained using Monostream DDF.

\begin{figure}
	\centering
	\includegraphics[width=0.7\columnwidth]{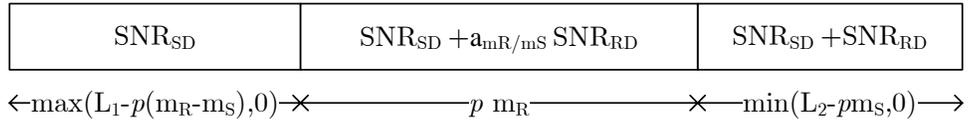}
	\caption{\label{fig:CanalSNR3blocs}Long-term SNR channel resulting after Patching $p$ time-slots of the second phase to generate $2^{m_R}$QAM symbols.}
\end{figure}

Assuming that the relay combines symbols in order to form $2^{m_R}$ QAM symbols during $p$ time-slots of the second phase, the resulting long term SNR channel is presented in \reffig{fig:CanalSNR3blocs}. It is composed of three blocks, the first one linked to $SNR_{SD}$ containing $max(L_1-p(m_R-m_S),0)$ bits, the second block is composed of the $pm_R$ bits combined by Patching at the destination and thus experiencing a long term SNR equal to $SNR_{SD}+a _{m_R/m_S}SNR_{RD}$. The last block is composed of the $\min(L_2-pm_S,0)$ bits transmitted on both links without Patching and thus experiencing a long term SNR equal to $SNR_{SD}+SNR_{RD}$.
This channel is different from the one obtained without Patching (\reffig{fig:nSNRchannel}). But, as the blocks characterized by $SNR_{SD}+a _{m_R/m_S}SNR_{RD}$ and $SNR_{SD}+SNR_{RD}$ have the same macro diversity behaviour, they are part of the same SNR block according to the definition of the Matryoshka channel. Thus, when the relay patches $p$ time-slots of the second phase, the resulting long-term SNR channel is a Matryoshka channel $\mathcal {M}((2,1),(L'_2 , L'_1))$ with $L_2'=pm_R+ \min(L_2-pm_S,0)$ and $L_1'=max(L_1-p(m_R-m_S),0)$.
Consequently, considering a fixed frame length, the relay can adapt its modulation so that the remaining block of diversity 2 contains at least $K$ bits, $L_2' \geq K$, and thus making full macro diversity achievable.
Note that, instead of considering the whole frame length in order to decide if Patching should be used, the relay could decide to guaranty that the full macro diversity is achievable after receiving half of the frame for improving the spectral efficiency of closed-loop systems.

Unfortunately, the use of bigger constellation resulting from the construction of hyper-symbols $z_k$ involves a coding gain loss.
Consequently, the relay can optimize the performance by using our proposed Patching technique with Minimal Use (MU) in which the number of hyper-symbols and the spectral efficiencies are minimized under the constrain of full macro diversity.
A generic optimization is out of the scope of this paper, but particular examples are provided in \refsec{Part:Perf}.

Note that our proposed Patching technique cannot be easily extended to the multiple-relay case.
\section{Micro diversity orders of DDF relaying schemes}\label{Part:MicroDiv}

In the previous section, we have designed practical DDF protocols that exploit the macro diversity of the system. In this section, we further improve the system performance by exploiting the micro diversity for the single-relay case.

\subsection{Links between micro and macro diversity}
Given an instant of correct decoding at the relay, for all relaying schemes whose equivalent fading channel can be modeled by a block fading channel or a Matryoshka channel, as soon as a micro diversity order $d$ is achievable, a macro diversity order of at least $d/N_r$ is achievable.
Indeed, the micro diversity is the minimal number of independent fading variables adding coherently in the performance expression and each fading coefficient is weighted by a long term SNR.

\subsection{Micro diversity behavior of several Monostream DDF schemes}
\subsubsection{Micro diversity behavior of the Monostream DDF scheme}
The block fading channel created by the use of the Monostream DDF scheme is composed of two blocks with fadings $h_{SD}\sqrt{P_S}$ and $h_{SD}\sqrt{P_S}+h_{RD}\sqrt{P_R}$, respectively. Each resulting fading coefficient carries a diversity order of one. It is shown in \cite{gresset2010mmp} that the micro diversity  of this channel under discrete input constraint is equal to the one of a block fading channel with independent fadings.
The upper bound on the diversity order of a coded modulation transmitted on a block-fading channel of equal length blocks has been derived in \cite{knopp2000cbf}. 
Due to the dynamic decoding time at the relay, we propose in \cite{plainchault2010ddfrus} a generalization of the Singleton bound on the diversity order to unequal length block-fading channels.
It results from these bounds that the monostream DDF scheme achieves full micro diversity if both blocks contain at least $K$ bits.

Consequently, the micro diversity order of the monostream DDF scheme is constrained by a block fading channel while the macro diversity order is constrained by a Matryoshka channel.
Thus, in the single-relay case, the block of highest diversity must contain at least $K$ bits ($L_2 \geq K$) in order to guarantee full macro diversity whereas each block must contain at least $K$ bits ($\min(L_1,L_2) \geq K$) in order to guarantee full micro diversity.

\subsubsection{Improvement of the micro diversity order by using Monostream DDF with modulation adaptation}
Channel coding and the frame subdivision is designed in order for the relay to  have access to enough information after receiving the first sub-frame to correctly decode the message. Thus, the condition $L_1 \geq T_1m_S\geq K$ is always satisfied.
The condition to be satisfied in order to guarantee the full micro diversity order at the destination thus becomes $L_2 \geq K$.
The modulation cardinality can be chosen accordingly by the source and the relay during the second phase of the protocol to reach full micro diversity.

\subsubsection{Improvement of the micro diversity order by using Patched Monostream DDF}
When using the Patched Monostream DDF relaying scheme, some symbols transmitted by the source during the first phase are patched so that, in the equivalent channel after patching, they are part of the block of highest macro diversity order. Consequently, after Patching, a lower number of symbols are part of the first channel block: increasing $L'_2$ involves decreasing $L'_1$.
Consequently, if $L'_1\leq K$, the Patching technique is useless for improving the micro diversity order.
We will see in the following how to use distributed space-time block code for always guarantying both the macro and micro diversity exploitation.

\subsection{Micro diversity behavior of several Distributed Alamouti (DA) DDF schemes}
\subsubsection{Micro diversity behavior of DA DDF scheme}
it has been proposed in \cite{laneman2002distributed} to use Distributed Space-Time Block Codes (DSTBC) in order to recover the spatial diversity offered by the virtual MIMO scheme formed in the relay channel.

The DA DDF relaying scheme is proposed in \cite{murugan2006clc} and is shown to achieve the diversity multiplexing tradeoff of the DDF protocol.
Using this relaying scheme, after its correct decoding, the relay builds all symbols, $\{ x_{m+1}, x_{m+2},\cdots \}$ with $m = L_1/m_S$ concurrently sent by the source. Thus, the relay transmits
\begin{equation}
	\left\{ \begin{array}{cc}
						-x_{m+k+1}^{*}\quad &\textrm{ if } k \textrm{ is odd}\\
						x_{m+k-1}^{*}\quad & \textrm{ if } k \textrm{ is even}
					\end{array} \right.
\end{equation}
where $k$ denotes the time-slot index of the second phase of the DDF protocol. As a remark, this protocol also allows the source to be relay-unaware.

After the classical Alamouti receiver, the resulting fading coefficient is $\mid \mat h_{SD}\mid ^{2}P_S + \mid \mat h_{RD}\mid ^{2}P_R$.
Thus, the transmitted frame's equivalent fading channel is a Matryoshka channel $\mathcal {M}((2N_r,N_r),(L_2 , L_1))$.
The condition that should be satisfied in order to guarantee full micro diversity is $L_2 \geq K$, which also involves  full macro diversity.
This condition only depends on the correct decoding time at the relay. 

\subsubsection{Improvement of the achievable micro diversity order using DA DDF with modulation adaptation}
As for the Monostream DDF scheme, the condition $L_2 \geq K$ can be met by allowing a signaling between the relay and the source and choosing the modulation cardinality accordingly.

\subsubsection{Improvement of the achievable micro diversity order using Patched DA DDF}
The Patching technique  combined with the Alamouti space-time code \cite{Alamouti1998} allows to keep the advantage of the low decoding complexity and a relay-unaware source.
Considering $2m_R/m_S$ time-slots, the destination receives the signals:
\begin{eqnarray}
 \mat{y}_{1,i} & = & x_{1,i}\sqrt{P_S}\mat{h}_{SD}+\mat{b}_{1,i},\quad 1 \leq i \leq 2(m_R/m_S-1)\\
 \mat{y}_{2,j}  & = & x_{2,j}\sqrt{P_S}\mat{h}_{SD}+z_{R,j}\sqrt{P_R}\mat{h}_{RD}+\mat{b}_{2,j},\quad j\in \{1,2\}.
\end{eqnarray}
During the two time-slots of the second phase of the DDF protocol, if the source transmits QPSK symbols, the relay transmits a combination of symbols forming $2^{m_R}$-QAM symbols:
\begin{eqnarray}
z_{R,1} &=&\sum_{k=1}^{m_R/m_S-1}a_k^{*}x^{*}_{1,k}+a_{m_R/m_S}^{*}x^{*}_{2,2},\quad a_k = \sqrt{\frac{3}{2^{m_R}-1}}2^{k-1}\\
z_{R,2} &=& -\sum_{k=1}^{m_R/m_S-1}a_k^{*}x_{1,k+m_R/m_S-1}^{*}-a_{m_R/m_S}^{*}x_{2,1}^{*},\quad a_k = \sqrt{\frac{3}{2^{m_R}-1}}2^{k-1}.
\end{eqnarray}
In order to build the Alamouti codeword, the destination makes the following signals combination:
\begin{eqnarray}
\tilde{\mat{y}}_{2,1}  & = &\sum_{k=1}^{m_R/m_S-1}a_k\mat y_{1,k+m_R/m_S-1}+a_{m_R/m_S}\mat y_{2,1},\quad a_k = \sqrt{\frac{3}{2^{m_R}-1}}2^{k-1}\\
\tilde{\mat{y}}_{2,2}  & = &\sum_{k=1}^{m_R/m_S-1}a_k\mat y_{1,k}+a_{m_R/m_S}\mat y_{2,2},\quad a_k = \sqrt{\frac{3}{2^{m_R}-1}}2^{k-1}.
\end{eqnarray}
After this combination, the resulting fading channel and SNR channel are Matryoshka channels \\$\mathcal {M}((2N_r,N_r),(L'_2 , L'_1))$ and $\mathcal {M}((2,1),(L'_2 , L'_1))$ respectively, with $L'_1 = max(L1-L_2(\frac{m_R}{m_S}-1), 0)$ and $L'_2 = min(L_1+L_2, L_2\frac{m_R}{m_S})$.
The relay adapts its modulation to satisfy : $m_R \geq \frac{Km_S}{L_2}$ in order to recover full micro diversity and by inference full macro diversity.

\subsection{Improvement of the achievable micro diversity order using Patched DSTBC DDF schemes}
We propose two other Patched DSTBC schemes \cite{plainchault2010pdstbc} so as to improve the micro diversity order:  the Patched Golden Code \cite{belfiore2005golden} and the Patched Silver Code \cite{hollanti2008Silver}.
As for the Patched Monostream and Patched Alamouti, these Patched DSTBCs enable an adaptative choice of the modulation by the relay in order to guarantee full diversity and these enable the source to be relay-unaware. 

These two Patched DSTBCs are used following the same scheme. Let us consider the first two time slots of the second phase and $2(m_R/m_S)$ time-slots from the first phase of the DDF protocol.
First, the relay generates 2 symbols from a $2^{m_R}$ modulation combining symbols transmitted during the first phase of the DDF protocol:
\begin{eqnarray}
z_{1} &=& \sum_{k=1}^{m_R/m_S} a_kx_{1,k} \quad \textrm{with } a_k=\sqrt{\frac{3}{2^{m_R}-1}}2^{k-1}\\
z_{2} &=& \sum_{k=m_R/m_S+1}^{2(m_R/m_S)} a_{k-m_R/m_S}x_{1,k}\quad \textrm{with } a_j=\sqrt{\frac{3}{2^{m_R}-1}}2^{j-1}
\end{eqnarray}
Then the relay transmits during the first time-slot of the second phase $z_{R,1} = f_1(z_{1},z_{2}, x_{2,1}, x_{2,2})$ and $z_{R,2} = f_2(z_{1},z_{2}, x_{2,1}, x_{2,2})$ during the second time-slot. These functions depend on the space-time code to be generated after combination at the destination side. They are given in \reftab{tab:FctTransR}.

The second step of the Patching technique is done at the destination side. It first builds the received signals ($\mat{\tilde{y}}_{1,1}$, $\mat{\tilde{y}}_{1,2}$) corresponding to the emission of $z_{1}$ and $z_{2}$ by the source:
\begin{eqnarray}
\mat{\tilde{y}}_{1,1} &=& \sum_{k=1}^{m_R/m_S} a_k \mat y_{1,k} \quad \textrm{with } a_k=\sqrt{\frac{3}{2^{m_R}-1}}2^{k-1}\\
\mat{\tilde{y}}_{1,2} &=& \sum_{k=m_R/m_S+1}^{2m_R/m_S} a_{k-m_R/m_S+1} \mat y_{1,k}\quad \textrm{with } a_j=\sqrt{\frac{3}{2^{m_R}-1}}2^{j-1} 
\end{eqnarray}
The destination then realizes a combination $\mat Y $ of the signals $\mat {\tilde{Y}}_1 = \left[\begin{array}{cc}\mat{\tilde{y}}_{1,1} & \mat{\tilde{y}}_{1,2}\end{array} \right]$ and $\mat {Y}_2 = \left[\begin{array}{cc}\mat y_{2,1} & \mat y_{2,2} \end{array}\right]$, in order to generate the desired space-time codewords. These combinations are given in \reftab{tab:PatchedDSTBDCombi} for each considered space-time code. They result in a signal of the form
\begin{equation}
\mat Y = c \left[\begin{array}{cc}\mat h_{SD} & \mat h_{RD}\end{array} \right ] \mat X_{STBC} + \mat B
\end{equation}
in which $c$ is a constant, $\mat X_{STBC}$ is the generated codeword of the considered space-time code and $\mat B$ is a matrix composed of Gaussian noise samples of zero mean and unit variance.

Consequently, using Patched Golden code or Patched Silver code, $2(m_R+m_S)$ bits are transmitted for each two time-slots of the second phase.  The resulting fading channel is a Matryoshka channel $\mathcal {M}((2N_r,N_r),(L'_2 , L'_1))$ with $L'_1= max(0, L_1-\frac{L_2}{m_S}m_R)$ and $L'_2 = min(L_2+L_1,\frac{L_2}{m_S}(m_R+m_S))$.
The relay adapts its modulation to satisfy : $m_R \geq m_S(\frac{K}{L_2}-1)$ in order to recover full micro diversity and by inference full macro diversity.

As for classical MIMO schemes, the Patched Golden code and Patched Silver code has a higher coding gain when compared with the Patched Alamouti, at the price of an increase decoding complexity. Because we are particularly interested in low complexity decoders, in the following, results are only given for the Patched Alamouti case.

\section{Simulation Results}\label{Part:Perf}
All presented performance are generated using mutual information computation with finite symbol alphabet whose performance can be achieved using specific modulation and coding schemes \cite{nguyen2007tlb}.
These performance are all given for the single-relay case.

\begin{figure}
	\centering
	\includegraphics[width=0.55\columnwidth]{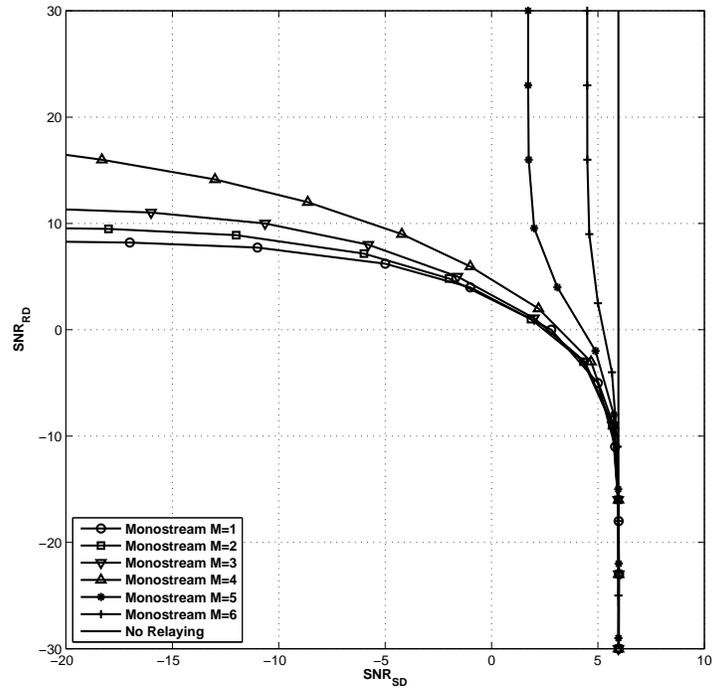}
	\caption{\label{fig:PoutMacroMono}Couples of SNRs achieving an outage probability of $10^{-2}$ up to the instant of correct decoding at the relay, considering a frame composed of 7 sub-frames, the first one being three time longer than the others and containing only information bits. The destination carries 2 antennas, the relay uses Monostream DDF.}
\end{figure}

\begin{figure}
	\centering
	\includegraphics[width=0.55\columnwidth]{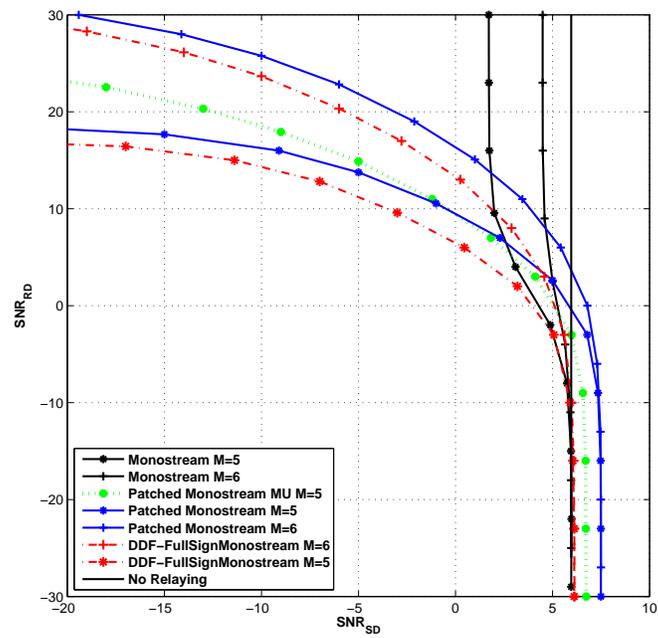}
	\caption{\label{fig:PoutMacroAllMono}Couples of SNRs achieving an outage probability of $10^{-2}$ up to the instant of correct decoding at the relay, considering a frame composed of 7 sub-frames, the first one being three time longer than the others and containing only information bits. The destination carries 2 antennas, distinct schemes are used at the relay.}
\end{figure}

\begin{figure}
			\centering				\includegraphics[width=0.55\columnwidth]{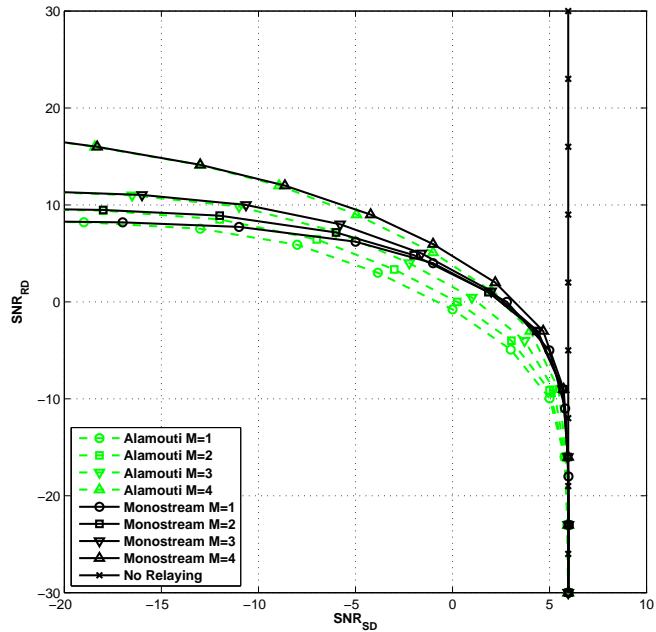}
			\caption{\label{fig:Micro_Pout_M1234} Comparison of the Monostream and Alamouti relaying scheme for different activation time of the relay in DDF protocols for $M= (1,2,3,4)$. $N_r$ = 2, $m_S=2$, $T_1= K/m_S$, and $T_1/T_{i, i \geq 1}=3$, Target outage probability of $10^{-2}$.}
\end{figure}

\begin{figure}
			\centering				\includegraphics[width=0.55\columnwidth]{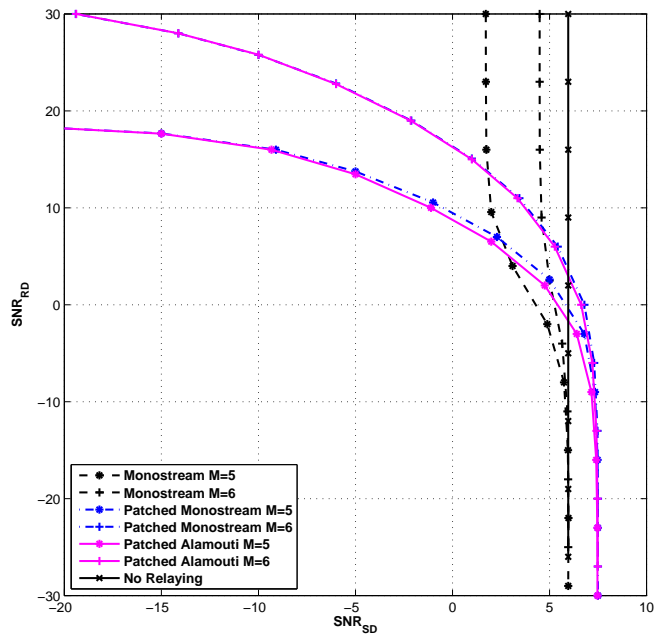}
			\caption{\label{fig:Micro_Pout_M56} Comparison of the Monostream and Alamouti relaying scheme for different activation time of the relay in DDF and Patched DDF protocols for $M=5,6$. $N_r$ = 2, $m_S=2$, $T_1= K/m_S$, and $T_1/T_{i, i \geq 1}=3$, Target outage probability of $10^{-2}$.}
\end{figure}

Considering the open-loop transmission mode, the couples ($SNR_{SD}$,$SNR_{RD}$) that allows to achieve the target outage probability of $10^{-2}$ are illustrated in \reffig{fig:PoutMacroMono}-\reffig{fig:Micro_Pout_M56} as a function of the block index $M$ after which the relay correctly decodes the message.  We consider a codeword composed of 7 subframes, the first one being three time longer than the others and containing only information bits. The source transmits QPSK symbols and the destination carries two reception antennas.

In \reffig{fig:PoutMacroMono}, the relay transmits by using a Monostream DDF scheme.
If the correct decoding of the relay occurs before the $4$-th sub-frame, the resulting Matryoshka channels using Monostream DDF guarantee that the full macro diversity is achievable: they guarantee that $L_2 \geq K$. This is illustrated by the fact that even if the $SNR_{SD}$ is low, it exists a $SNR_{RD}$ achieving the required quality of service of $10^{-2}$ (horizontal asymptote). When $M=5$ or $M=6$, the SNR channel block of highest diversity order contains $2K/3$ and $K/3$ bits, respectively. In both cases, full macro diversity is not achievable and a minimal value of $SNR_{SD}$ is needed to reach the target outage probability (vertical asymptote). 

In \reffig{fig:PoutMacroAllMono}, the relay transmits using a Monostream DDF scheme, Patched Monostream DDF, Patched Monostream with MU, or Monostream with modulation adaptation. We are interested in cases in which full macro diversity is not achievable with Monostream DDF ($M=5$ or $M=6$).
Using Patched Monostream relaying scheme, the relay sends combinations of symbols till the transmission end ($L_2' \geq K$).
In the Patched Monostream with MU case, the relay adapts the number of Patched symbols so that $L_2' = K$.
Note that, for some frames configuration, Patched DDF and Patched DDF with MU result in the same strategy and thus achieve the same performance.
If the relay correctly decodes after receiving the $5$-th sub-frame, the resulting SNR channel with Monostream is a $\mathcal {M}((2,1),(2K/3,7K/3))$ and the full macro diversity is not achievable, which is illustrated by the vertical asymptote; whereas using Patched Monostream in which the relay forms symbols of a 16-QAM, the resulting SNR channel is  a $\mathcal {M}((2,1),(4K/3,5K/3))$ thus the full macro diversity is achievable, which is illustrated by the horizontal asymptote.
If the relay uses Patched Monostream with MU, the resulting SNR channel is a $\mathcal {M}((2,1),(K,2K))$.
Thus, less 16 QAM symbols are transmitted guaranteeing a smaller coding gain loss which can be observed for low $SNR_{RD}$ when compared with Patched Monostream DDF.
If the relay correctly decodes after the $6$-th sub-frame, the resulting SNR channel with Monostream is a $\mathcal {M}((2,1),(K/3,8K/3))$, the full macro diversity is not achievable, whereas using Patched Monostream in which the relay forms symbols of a 64-QAM, the resulting SNR channel is  a $\mathcal {M}((2,1),(K,2K))$ thus the full macro diversity is achievable.
Comparing with Monostream DDF with modulation adaptation, full macro diversity can be recovered for $M=5$ and $M=6$, and there is no coding gain loss anymore as the source also transmits symbols from higher constellation.
Consequently, the coding gain loss is a price to pay for achieving a low signalling constrain but occurs at very low $SNR_{RD}$, i.e. cases in which the relay is useless. Furthermore, the relay can also select to transmit only for $SNR_{RD}$ values where a gain is brought by relaying.

In \reffig{fig:Micro_Pout_M1234}, the relay uses Monostream DDF or DA DDF and decoding events at the relay guaranteeing full macro diversity ($1 \leq M \leq 4)$ are considered. The gain achieved by DA DDF is for $SNR_{RD}$ close to $SNR_{SD}$. This gain represents the gain brought by the exploitation of the micro diversity as DA DDF brings micro diversity during the second phase of the protocol which is not the case of the Monostream DDF.

In \reffig{fig:Micro_Pout_M56}, the relay can use Patched Monostream DDF or Patched DA DDF.
When $M = 5$, and $M = 6$, using the Patching technique forming 16-QAM, and 64-QAM respectively, enables an horizontal asymptote for low $SNR_{SD}$, illustrating the full macro diversity behavior.

\begin{figure}
	\centering	\includegraphics[width=0.5\columnwidth]{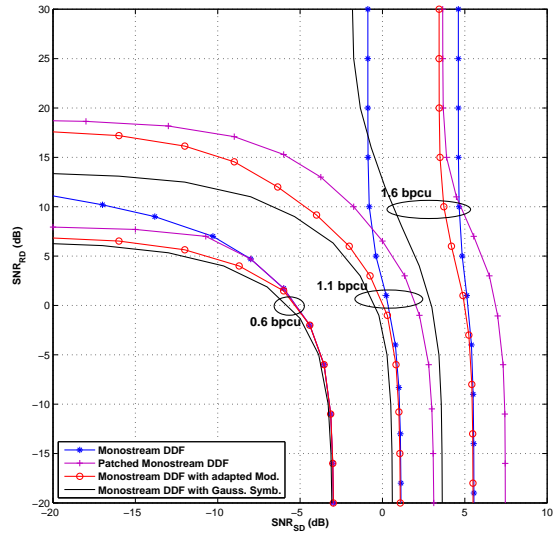}
	\caption{\label{fig:SLA-SpecEff-allMono}Couples of SNRs achieving different spectral efficiency are plotted up to the considered relaying schemes. The transmission occurs using HARQ and considering a frame composed of 3 sub-frames, the first one being four time longer than the others. Slow link adaptation is realized over the coding rates. $m_S = 2$, $N_r = 2$, $SNR_{SR}=10dB$.}
\end{figure}

\begin{figure}
	\centering	\includegraphics[width=0.5\columnwidth]{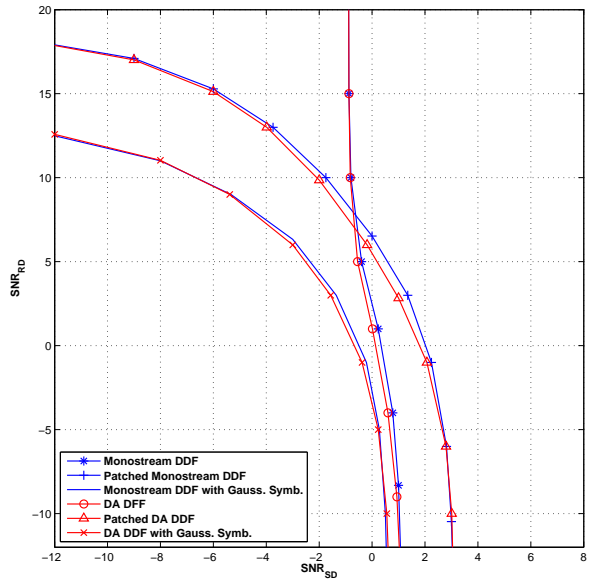}
	\caption{\label{fig:SLA-SpecEff-11bpcu_AlamVSMono}Couples of SNRs achieving a spectral efficiency of 1.1 bpcu, using HARQ and considering a frame composed of 3 sub-frames, the first one being four time longer than the others. Slow link adaptation is realized over the coding rates. $m_S = 2$, $N_r = 2$, $SNR_{SR}=10dB$.}
\end{figure}

For the closed loop transmission mode, we consider a transmission using HARQ and slow link adaptation: the spectral efficiency is maximized over the available coding rates (0.5, 0.6, 0.7, 0.8, 0.9 or 1 at the end of the first sub-frame) in \reffig{fig:SLA-SpecEff-allMono} and \reffig{fig:SLA-SpecEff-11bpcu_AlamVSMono}.
In these figures, the considered codeword is composed of maximum 3 sub-frames, the first one being 4 time longer than the others. The source transmits QPSK symbols, the destination carries 2 reception antennas and $SNR_{SR} = 10dB$.

In \reffig{fig:SLA-SpecEff-allMono}, the couples of $SNR_{SD}$ and $SNR_{RD}$ achieving distinct target values of spectral efficiency (0.6, 1.1 and 1.6 bpcu) are plotted for the Monostream DDF, the Patched Monostream DDF with MU, the Monostream DDF with adapted modulation and the performance achieved using Monostream DDF with a Gaussian symbol alphabet.
The performance achieved with a Gaussian symbol alphabet are the best performance achievable using Monostream scheme. For low $SNR_{SD}$, the Patched Monostream DDF protocol enables to fill the gap between the performance achieved using Monostream DDF and the best achievable performance. But for low $SNR_{RD}$, performance using Monostream DDF are better because the Patching technique introduce a coding gain loss due to the generation of higher modulation. A selection of the best scheme can be done according to the observed SNRs, and a slow link adaptation can be done on the coding rate and on the relaying scheme.
The performance achieved with Monostream DDF with adapted modulation enables to fill the gap between Patched Monostream DDF and Monostream DDF with Gaussian symbol alphabet.
Consequently, if signalling can be afford, as for the downlink of a cellular transmission, the performance are maximized using Monostream DDF with adapted modulation. If signalling between source and relay is costly, as for the uplink of a cellular transmission, the performance are maximized using Patched Monostream DDF.

Note that because HARQ is used in these different figures, the maximal achievable spectral efficiency never can be achieved through the relay-destination link only. Indeed, the maximal spectral efficiency requires the destination to correctly decode the message after receiving the first sub-frame which can not be transmitted by a causal relay.

In \reffig{fig:SLA-SpecEff-11bpcu_AlamVSMono}, the couples of $SNR_{SD}$ and $SNR_{RD}$ achieving the target spectral efficiency 1.1 bpcu are plotted for the Monostream DDF relaying scheme and DA DDF relaying scheme. The gain brought by the DA DDF scheme is noticeable when the SNRs of the source-destination link and relay-destination link are close. Thus, an important gain can be brought only taking into account the achievable macro diversity order prior to the micro diversity order.
The macro diversity order is thus a relevant metric for the relay channel.

\section{Conclusion}\label{Part:Conclusion}
In this article, we propose practical implementation of the DDF protocol, already known to achieve good Diversity Multiplexing tradeoff, with channel coding. We consider both cases in which signalling is allowed between the source and the relay, and in which the source is relay-unaware.

Upper bounds for finite symbol alphabet on the achievable macro diversity order have been derived for the Monostream DDF and the DA-DDF relaying schemes, the adapted modulation DDF schemes and the Patched DSTBC DDF schemes.

In an open loop transmission, the Patching technique improve the outage probability by increasing the achievable macro diversity order.
In the closed-loop transmission mode, we have shown that the Patching technique enables to fill the gap of spectral efficiency between a relaying scheme not achieving full macro diversity and the best theoretical performance obtained with a Gaussian symbol alphabet.
As Patching technique is particularly efficient to improve the outage probability and the spectral efficiency of the studied relaying protocol, it would be of interest to adapt this concept to other channels such as the interference channel with relay and the multiple access channel with relay.

\newpage

\newpage

\begin{table}
\centering
\begin{tabular}{|c|c|c|}
\hline
 Patched DSTBC    & $z_{R,1} $ 											 &	 $z_{R,2}$												\\ \hline
 Patched Golden   & $i\frac{\overline{\phi}}{\phi}(x_{2,2}+\overline{\alpha}z_{2})$ with $\left \{ 				\begin{array}{ccl}\alpha &=& \frac{1+\sqrt{5}}{2}\\
 									\overline{\alpha} &=& \frac{1-\sqrt{5}}{2}\\
 									\phi &=& 1+i-i\alpha \\
 									\overline{\phi} &=&1+i-i\overline{\alpha}\\
 									i^2 &=& -1 \end{array} \right .$
 									   & $\frac{\overline{\phi}}{\phi}(x_{2,1}+\overline{\alpha} z_{1})$  with $\left \{ 				\begin{array}{ccl}\alpha &=& \frac{1+\sqrt{5}}{2}\\
 									\overline{\alpha} &=& \frac{1-\sqrt{5}}{2}\\
 									\phi &=& 1+i-i\alpha \\
 									\overline{\phi} &=&1+i-i\overline{\alpha}\\
 									i^2 &=& -1 \end{array} \right .$\\ \hline
 Patched Silver   & $-x_{2,2}^*-\frac{(1-2i)z_{1}^*+(1+i)z_{2}^*}{\sqrt{7}}$ \quad with $i^2 = -1 $& $ x_{2,1}^*+\frac{(-1+i)z_{1}^*+(1+2i)z_{2}^*}{\sqrt{7}}\quad $with $i^2 = -1 $								\\ \hline
\end{tabular}
\caption{\label{tab:FctTransR}Symbols to be transmitted by the relay during the two time-slots of the second phase according to the considered Patched DSTBC.}
\end{table}

\begin{table}
\centering
\begin{tabular}{|c|c|c|}
\hline
Patched DSTBC    & $\mat Y$ & Value of the constant $c$\\ \hline
Patched Golden   & $\frac{\phi ( \alpha \mat{\tilde{Y}}_1 + \mat Y_2)}{\sqrt{|\phi^2 (1+\alpha^2)|}}$ & $ \frac{1}{\sqrt{2}}$\\ \hline
Patched Silver   & $\frac{\mat {\tilde{Y}_1}}{\sqrt{7}} \left[ \begin{array}{cc}
 																1+i & -1+2i \\ 
 																-(1+2i) & (-1+i) \\
 															 \end{array} \right] +\mat Y_2$ & $\frac{1}{\sqrt{2}}$\\
\hline
\end{tabular}
\caption{\label{tab:PatchedDSTBDCombi}Combinations done at the destination in order to build the space-time codewords according to the considered Patched DSTBC.}
\end{table}

\newpage

\end{document}